# Strain Switching in van der Waals Heterostructures triggered by a Spin-Crossover Metal Organic Framework

*Carla Boix-Constant, Víctor García-López, Efrén Navarro-Moratalla, Miguel Clemente-León, José Luis Zafra, Juan Casado, Francisco Guinea, Samuel Mañas-Valero,\* Eugenio Coronado.\**

C. Boix-Constant, V. García-López, Dr. E. Navarro-Moratalla, Dr. M. Clemente-León, Dr. S. Mañas-Valero, Prof. E. Coronado.
Instituto de Ciencia Molecular, Universidad de Valencia, Catedrático José Beltrán 2, Paterna, 46980, Spain.
E-mail: samuel.manas@uv.es, eugenio.coronado@uv.es
Dr. J. Luis Zafra, Prof. J. Casado.
Department of Physical Chemistry, University of Málaga, Campus de Teatinos s/n, 229071 Málaga, Spain.
Prof. F. Guinea.
Instituto Madrileño de Estudios Avanzados en Nanociencia (IMDEA- Nanociencia), Calle Farady 9, Cantoblanco, Madrid 28049, Spain.

Van der Waals heterostructures (vdWHs) provide the possibility of engineering new materials with emergent functionalities that are not accessible in another way. These heterostructures are formed by assembling layers of different materials used as building blocks. Beyond inorganic two-dimensional (2D) crystals, layered molecular materials remain still rather unexplored, with only few examples regarding their isolation as atomically thin-layers. Here, we enlarge the family of van der Waals heterostructures by introducing a molecular building block able to produce strain: the so-called spin-crossover (SCO). In these metal-organic materials, a spin transition can be induced by applying external stimuli like light, temperature, pressure or an electric field. In particular, smart vdWHs are prepared in which the electronic and optical properties of the 2D material (graphene and WSe$_2$) are clearly switched by the strain concomitant to the spin transition. These molecular/inorganic vdWHs represent the deterministic incorporation of bistable molecular layers with other 2D crystals of interest in the emergent fields of straintronics and band engineering in low-dimensional materials.

# 1. Introduction

The spin-crossover (SCO) transition represents one of the most spectacular examples of molecular bistability.[1] Fe(II) compounds formed by octahedral metal sites provide archetypical examples of this phenomenon. In these, a transition from a paramagnetic high spin (HS) state with an open-shell configuration $-(t_{2g})^4(e_g)^2-$ to a diamagnetic low spin (LS) one with a closed-shell configuration $-(t_{2g})^6-$ (**Figure 1.a**).[2] Since in the HS state the $e_g$ orbitals – which have an antibonding character– are occupied, the average metal-ligand bond length is higher for the HS if compared with the LS, thus leading to a compression/expansion of the metal site and, consequently, of the crystal lattice upon the spin transition.[3–5]

Although the SCO phenomenon has been widely explored at the nanoscale,[6] its deterministic combination with other 2D materials has not been studied yet. So far, just graphene grown by chemical vapor deposition or liquid exfoliated $MoS_2$ nano-layers have been combined with sublimated or drop-casted SCO nanomaterials.[7–10] While these previous approaches benefit of its easy fabrication process and scalability in terms of performing devices, they lack of the conceptual simplicity of a vdWH, being not possible to change at will the stacking sequence of the different layers involved.[11] In fact, the interplay between the SCO and 2D materials is currently under debate, being its origin attributed to different effects as chemo-electric gating,[7] modulation of the doping level,[8] spin-dependent polarizability[9] or strain effects.[10]

As pointed above, the SCO phenomenon has been used to induce strain in 2D materials by linking in solution SCO nanoparticles on $MoS_2$ flakes.[10] Here, we introduce a novel technique for triggering strain on 2D materials by using thin SCO van der Waals layers. Current state-of-the-art regarding strain engineering on 2D materials include the thermal variation or bending of the substrate where the 2D material is located,[12–15] inducing ripples and bubbles,[16,17] applying hydrostatic pressure[18,19] or the use of piezoelectric and micro-heater actuators,[20–22] among others. Most of the previous approaches allow varying the degree of strain, although they are technically complex and not easy to integrate, for example, with low-temperature measurements. Using thin SCO layers for introducing strain on 2D materials is as easy as fabricating a van der Waals heterostructure. Although the amount of strain in our approach is fixed and determined by the employed SCO system, the wide versatility of coordination chemistry allow to tune the degree of strain (volume change upon the spin transition) and transition temperatures, offering a versatile toolbox of strain-inducing van der Waals materials.

Notice that so far 2D materials have been used as a tool for sensing the spin transition in these SCO compounds. In this work, we rather focus on the 2D crystal, studying the effect that this SCO compound has on its properties since, due to the insulating nature of the SCO, it is not expected that the 2D material modifies significantly the electronic properties of the molecular system. Nonetheless, we note that the fabrication of van der Waals heterostructures with graphene or other 2D materials can be employed for protecting these thin molecular layers upon, for example, laser irradiation. In particular, we explore how the electronic and optoelectronic properties of a 2D material are affected by the strain triggered by the spin transition. To reach this goal, we have prepared in a deterministic way vdWHs formed by SCO thin layers and inorganic 2D materials, such as semiconducting few-layers graphene (FLG), direct band-gap semiconducting $WSe_2$ monolayer, metallic $NbSe_2$ and insulating hexagonal boron nitride (h-BN). As SCO system we have chosen the layered {Fe-(3-Clpyridine)$_2$-[Pt$^{II}$(CN)$_4$]} metal-organic framework (**Figure 1.b**), that presents a change in volume of *ca.* 8 % upon the spin transition, with a thermal hysteresis centered at 150 K (see **Supplementary Section 6**).[23]

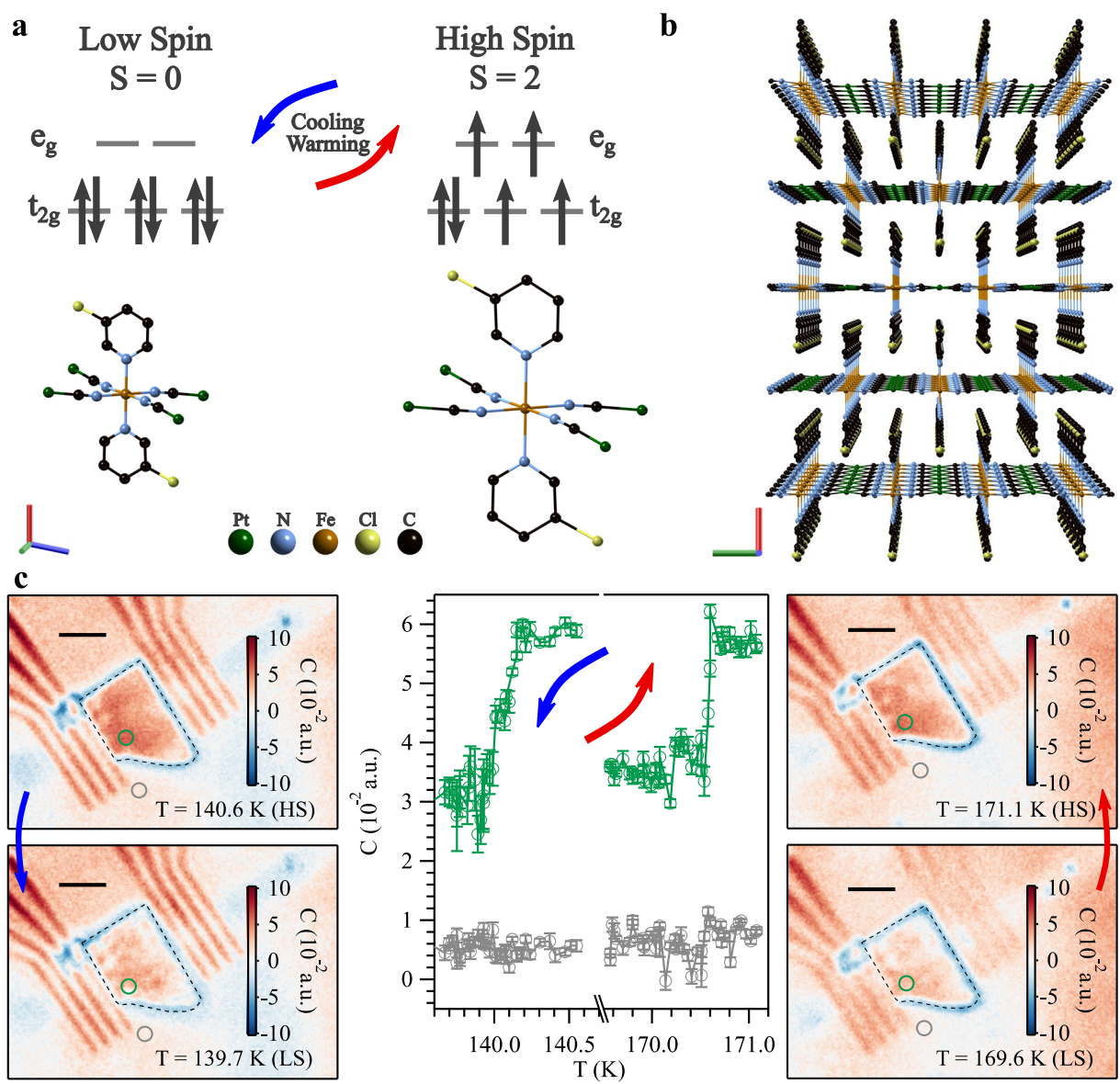

**Figure 1.-** Spin-crossover (SCO) transition in the layered molecular MOF {Fe-(3-Clpyridine)$_2$-[Pt$^{II}$(CN)$_4$]}. a) Scheme of the SCO phenomenon for a d$^6$ (Fe$^{2+}$) metal complex. For clarity, the structural change of the molecule associated with the SCO transition (*ca.* 8 % volume change, see **Supplementary Section 6**) has been exaggerated. The red, green and blue axis correspond to the a, b and c crystallographic directions, respectively. b) Detail of the layered structure of the SCO material. c) Optical contrast images (vdWH A.1 in the **Supplementary Information**) at different selected temperatures together with the contrast temperature dependence (central panel) for a mechanically exfoliated SCO thin-layer (enclosed with black dashed lines). The optical contrast depicted in the central panel is calculated in the positions marked with green (SCO) and grey (background) marks in the optical contrast images (see **Methods** for details). The blue (cooling) and red (warming) curves indicate the hysteresis direction. The high spin and low spin state is denoted as HS and LS, respectively. Scale bar: 5 μm.

## 2. Results and Discussion

The SCO compound Fe-(3-Clpyridine)$_2$-[Pt$^{II}$(CN)$_4$]} can be mechanically cleaved to afford thin crystals, as performed with graphene or other 2D molecular materials.[24–26] The thermal bistability of this process can be tracked by optical contrast (**Figure 1.c**) since the change in the lattice parameters concomitant to the spin transition results in a modification of both the thickness and the refractive index of the material.[23,27] Thus, a clear hysteresis loop can be observed (**Figure 1.c,** central panel and in the **Supplementary Movie**). Different SCO thin-layers are inspected (**Supplementary Section 3**), observing different spin-state propagation mechanisms. Typically, in the thickest SCO analyzed (~ 1 µm), different domains are created, with even the appearance of cracks in the first thermal cycle, while in the thinner ones (ca. 300 nm) we observe a single domain wall propagation, being these layers robust enough upon several thermal cycles. In addition, the spin transition can be also tracked by Raman spectroscopy, which shows an enhanced variation in the energy bands associated to the C≡N and Fe-N vibrational stretching modes within the spin transition temperatures (see **Supplementary Section 2).**

Using these thin SCO crystals we have prepared SCO/FLG vdWHs in which the FLG is on top of the SCO and investigate the electrical transport properties (**Figures 2.a-b**). By performing 4-probe DC electronic transport measurements, we have characterized simultaneously the SCO/FLG vdWH and, as a reference, the isolated FLG (see **Methods**). In **Figure 2.c** we plot the thermal dependence of the resistivity, ρ. Considering the vdWH, a clear hysteresis loop is observed, where the transition temperatures (T$_c$) are highlighted with blue – cooling– and red –warming– dashed lines. Notably, this hysteresis only occurs for the vdWH whereas it is absent for the FLG case. The T$_c$ values are in accordance with the reported temperatures for the bulk SCO compound, that are 141.4 K for cooling down and 164.5 K for warming up.[23] The hysteresis loops are robust over several thermal cycles for the different prepared devices (see **Supplementary Section 4**). Next, we have investigated the influence of a gate voltage in the electrical response of the vdWH (**Figure 2.d**). For addressing the influence of the SCO thin-layer, we compute the resistivity increment, Δρ(T, V$_g$), defined as Δρ (T, V$_g$) = ρ$_{warming}$(T, V$_g$) – ρ$_{cooling}$(T, V$_g$). As shown in **Figure 2.e**, the transition temperatures can be clearly observed for the vdWH, as a crossover between the blue region, where Δρ(T, V$_g$) = 0, and the red zone, where Δρ (T, V$_g$) ≠ 0. On the contrary, within the hysteresis loop Δρ(T, V$_g$) exhibits a gate dependence, with a minimum centered at around 20 V, coinciding with the charge neutrality point (CNT) of the FLG. Regarding the FLG, no appreciable dependence is observed, being Δρ(T, V$_g$) zero, within the experimental error, for all the temperature and voltage range studied.

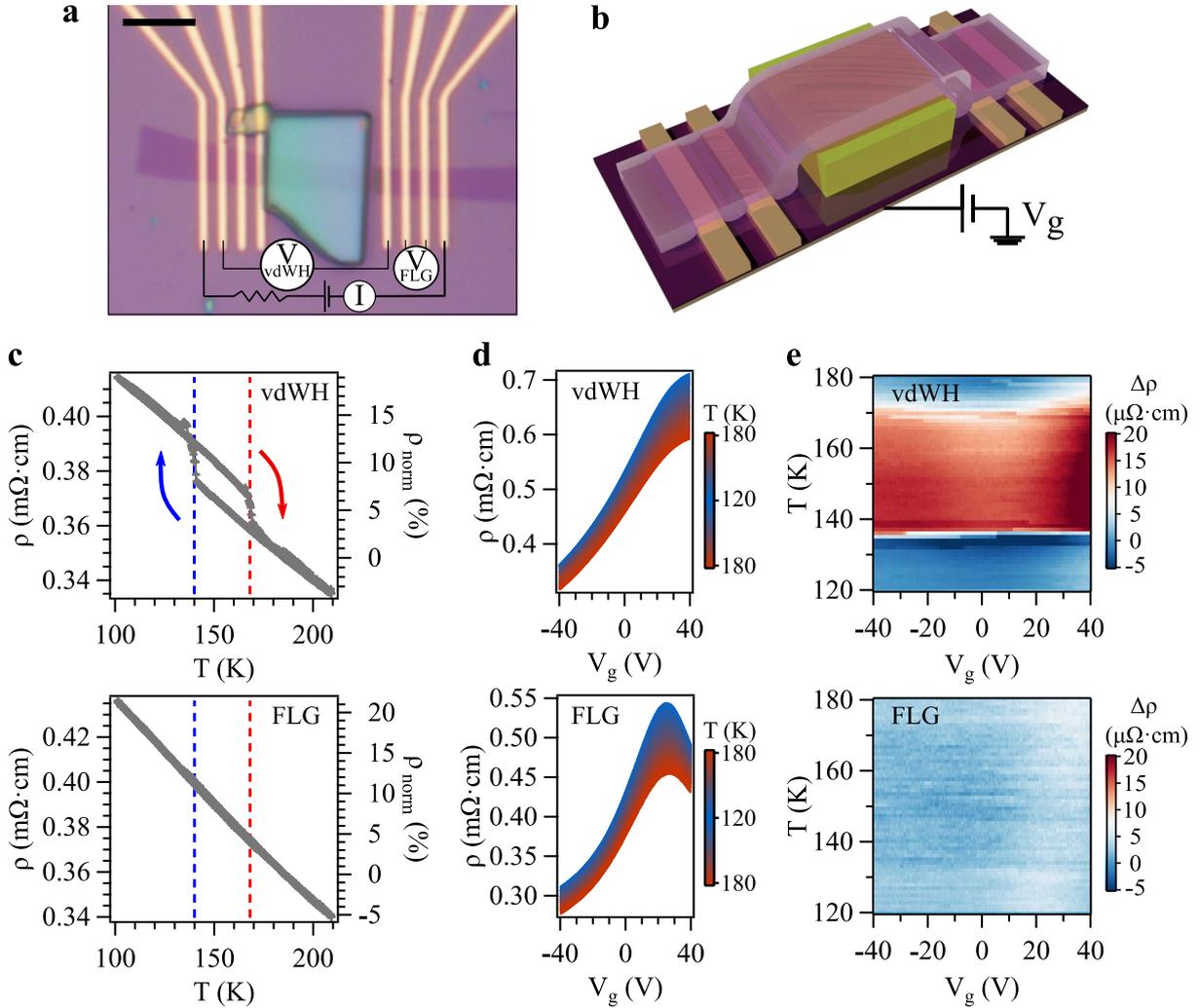

**Figure 2.-** Electrical characterization of a SCO/FLG vdWH (vdWH A.1 in the **Supplementary Information**). a) Device with the transport configuration sketched showing the central SCO thin-layer and FLG across it. Scale bar: 5 μm. b) Artistic view of the vdWH. The SCO is represented in yellow, the FLG in purple and the contacts pad in gold. The use of a back gate voltage is sketched. c) Thermal dependence of the resistivity for the vdWH and the FLG. The transition temperatures for cooling (warming) are marked as blue (red) dashed lines. The blue (cooling) and red (warming) arrows indicate the hysteresis direction. d) Gate voltage dependence of the resistivity for various temperatures. e) Resistivity increment, Δρ(T, V$_g$), where Δρ(T, V$_g$) = ρ$_{warming}$(T, V$_g$) − ρ$_{cooling}$(T, V$_g$), as a function of the gate voltage and temperature.

Thanks to the versatility offered by the vdWHs, different stacking configurations between the SCO and the 2D material are investigated. In particular, we consider the FLG placed above the SCO thin layer (in short SCO/FLG, vdWH type A) and the FLG placed below the SCO layer, in direct contact with the silicon substrate (FLG/SCO, vdWH type B). In addition, an intermediate h-BN layer is placed in between the FLG and the SCO layer (SCO/h-BN/FLG and FLG/h-BN/SCO for vdWH types C and D, respectively) in order to minimize proximity electronic effects present owing to the contact between FLG and SCO. A hysteresis transition is clearly observed for all the vdWHs although the relative change in resistivity, $\Delta\rho_{norm}$, within the hysteresis depends on the stacking configuration. The highest $\Delta\rho_{norm}$ values, up to 7 %, are obtained for the SCO/FLG (**Figure 3.a**), being reduced to a 2 % in the reverse configuration, *i.e.*, FLG/SCO (**Figure 3.b**). The same trend is observed when an insulating h-BN layer is decoupling the FLG and the SCO, reducing $\Delta\rho_{norm}$ from 1.5 % in SCO/h-BN/FLG (**Figure 3.c**) to 1.2 % in FLG/h-BN/SCO (**Figure 3.d**). In the same manner, we have placed a band metal, namely $NbSe_2$. In the $NbSe_2$-based vdWHs, the spin transition is not detected, regardless if the $NbSe_2$ thin-layer is placed above (SCO/$NbSe_2$, vdWH type E; **Figure 3.e**) or below ($NbSe_2$/SCO, vdWH type F; **Figure 3.f**) the SCO thin-layer. The complete set of thermal cycles, different devices and geometrical factors can be consulted in the **Supplementary Section 4**.

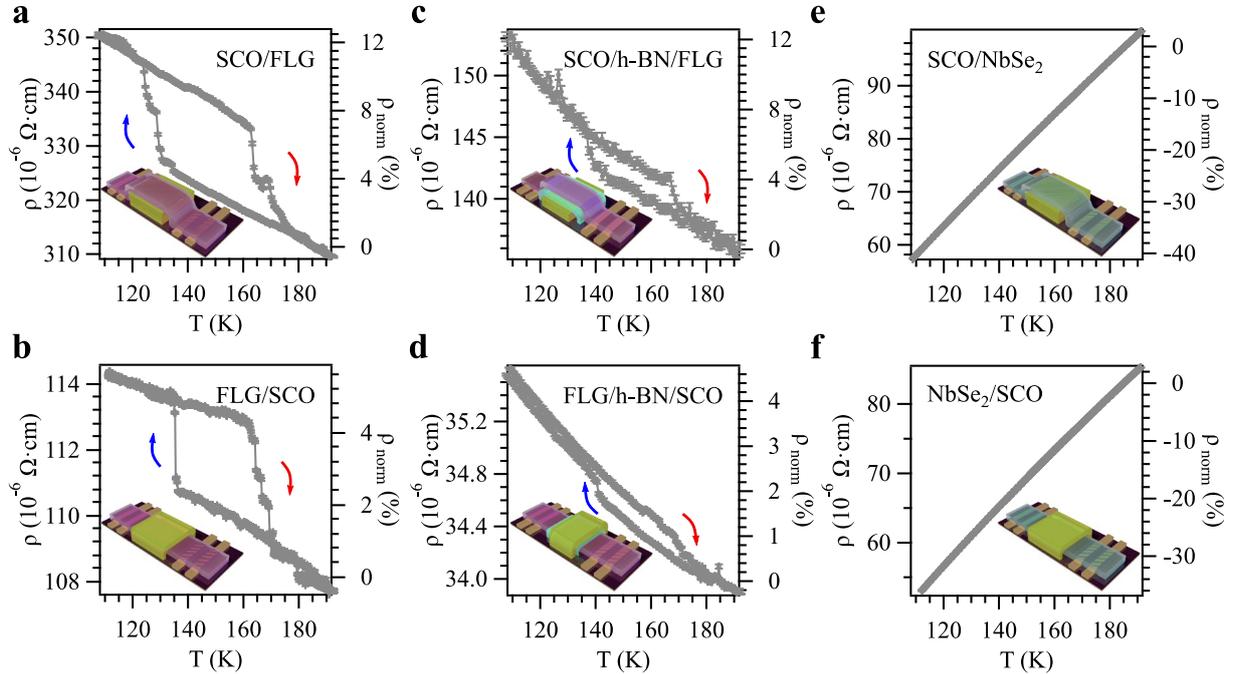

**Figure 3.** Thermal dependence of the resistivity for different vdWHs configurations. a) SCO/FLG (vdWH A.2 in the **Supplementary Information**). b) FLG/SCO (vdWH B.1 in the **Supplementary Information**). c) SCO/h-BN/FLG (vdWH C.1 in the **Supplementary Information**). d) FLG/h-BN/SCO (vdWH D.1 in the **Supplementary Information**). e) SCO/$NbSe_2$ (vdWH E.1 in the **Supplementary Information**). f) $NbSe_2$/SCO (vdWH F.1 in the
7

**Supplementary Information**). A schematic cartoon of the device is represented as an inset. The SCO, FLG, h-BN and NbSe$_2$ layers are represented in yellow, purple, bright-blue and greenish colors, respectively. The contacts are shown in gold color. The blue (cooling) and red (warming) arrows indicate the hysteresis direction.

Overall, the SCO transition is clearly detected in a reproducible and robust way by transport measurements in the FLG-based vdWHs, yielding an enhancement (diminishing) of the resistivity while sweeping from HS (LS) to LS (HS). The magnetic transition increases the resistivity of FLG by about Δρ/ρ ~10 % (**Figure 2**). A plausible explanation of this effect is the presence of strains in the FLG due to the compression or expansion induced by the SCO, yielding to an estimated mean free path $\ell_r \sim 120$ nm, with a height modulation of $h_r \sim 1.8$ Å, and typical strains $\bar{u}_{ij} \sim h_r^2/\ell_r^2 \leq 2$ %. We assume that the SCO induces, or increases already existing ripples in the FLG since many mechanisms which induce local compressions or dilatations lead to out-of-plane ripples. Although an electronic origin (produced by the change in the dielectric constant upon the spin transition, *i.e.*, attributed to the different polarizabilities in the LS and HS caused by the structural variations concomitant to the spin transition) cannot be fully discarded, a strain effect is estimated to modify the conductivity stronger since strains couple to electrons by modifying the carbon-carbon distances, which, in turn, change the hopping of electrons residing in $p_z$ orbitals in neighboring carbon atoms (see **Suplementary Section 4.5** for further details).[28] In addition, note that absolute values of the dielectric constant in SCO compounds are moderate, with a smaller change in those compounds undergoing the spin transition at lower temperatures.[29] The resistivity change is larger when the FLG is on top of the SCO. This behavior is compatible with the effect of an in-plane and out-of-plane compression (expansion) while cooling (warming) when the SCO is underneath (SCO/FLG), but only an in-plane compression (expansion) when the SCO is on top (FLG/SCO). The same tendency is observed if an h-BN layer is placed between the SCO and the FLG but it is absent in the case of using the metal NbSe$_2$. This observation supports as well that electronic interactions or proximity effects do not seem to be the main origin of the interaction. More likely, we consider the volume change associated with the spin transition, which produces an effective strain in the 2D material that shifts the Fermi level. In accordance, no substantial variation in the conductivity is observed in a band metal as NbSe$_2$ since the density of states does not vary drastically.

Interestingly, a measurable and hysteretic Hall resistance is observed when a SCO thin layer is placed above a FLG Hall bar (**Figure 4.a-c**; see **Methods** for fabrication details). This change upon the spin transition is also clearly appreciable when a gate voltage is applied (**Figure 4.d-e).** In the FLG case, a residual, but non-hysteretic response is also found, suggesting the existence of a small residual magnetic field of a few Oe. The observed hysteretic effect in the vdWH can be understood in terms of strains associated to the magnetic transition of the SCO layer. These strains act like effective magnetic fields[30] on the electrons in the FLG. This additional field is to be added or subtracted to the residual magnetic field (depending on the valley in the FLG band structure), thus modifying the Hall response. The effect is small, but not negligible, and it changes slightly in different hysteretic cycles (see **Supplementary Section 4.3**). These results suggest that the strain induced magnetic field has a small value, as it reflects strain gradients which most likely average to zero, and that these strains change each time the sample is taken through the magnetic transition.

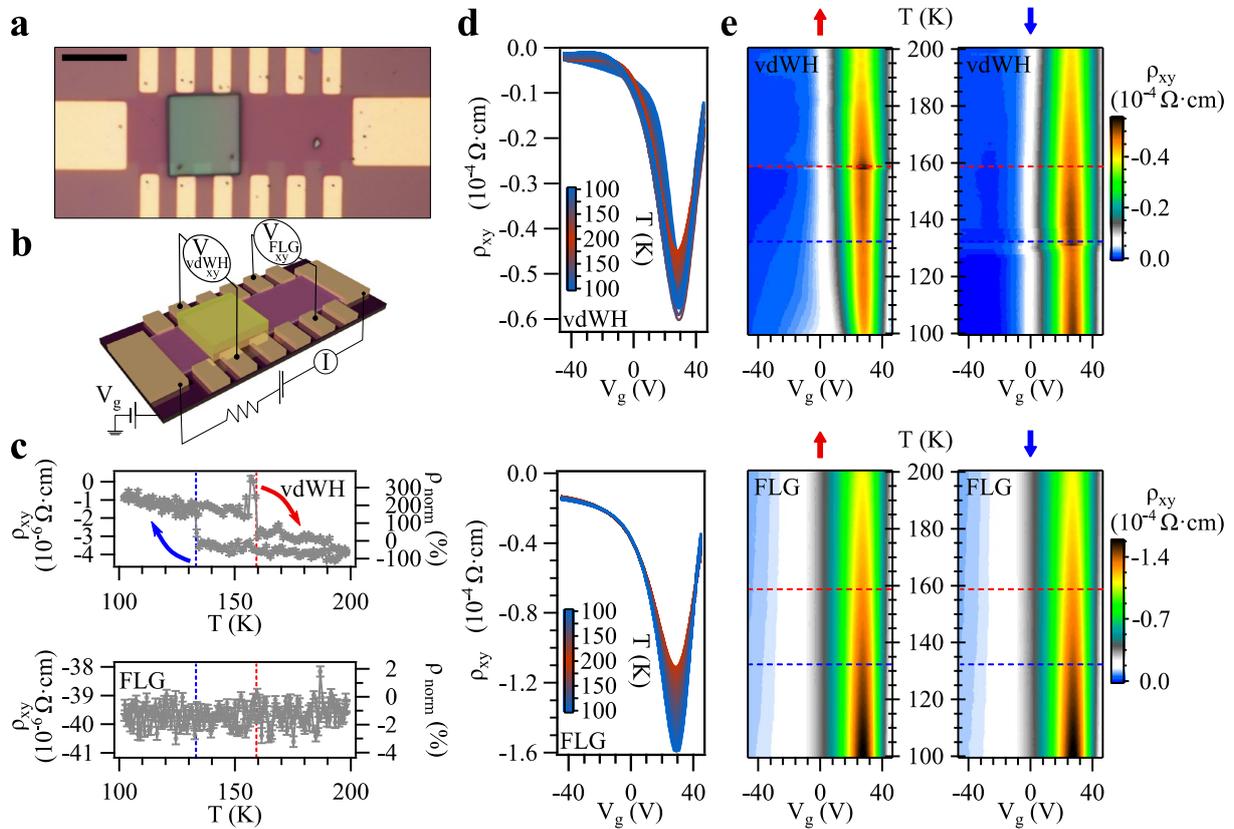

**Figure 4.-** Electrical characterization of a FLG/SCO vdWH (vdWH B.3 in the **Supplementary Information**). a) Device showing a FLG Hall bar with a SCO thin-layer placed on top. Scale bar: 5 μm. b) Artistic view of the vdWH with the transport configuration sketched. The SCO is represented in yellow, the FLG in purple and the contacts pad in gold. c) Fitted resistance from DC IV curves. The transition temperatures for cooling (warming) are marked as blue (red) dashed lines. The blue (cooling) and red (warming) arrows indicate the hysteresis direction. d) Gate voltage dependence of the resistivity for various temperatures with the corresponding 2D plot in e for warming and cooling cycles (indicated with a red and blue arrow, respectively).

Photoluminescence (PL) is another physical property dramatically modified by strain.[31] In **Figure 5.a-b,** a vdWH based on a WSe$_2$ monolayer placed below a SCO thin-layer (WSe$_2$/SCO, vdWH type G) is shown. The PL thermal dependence for a reference WSe$_2$ monolayer and the vdWH is presented in **Figure 5.c** and **Figure 5.d**, respectively. At 200 K, the PL of the reference WSe$_2$ monolayer and the vdWH are comparable, although the peak for the vdWH is slightly broader (see **Supplementary Section 5**). While cooling down, the PL peak narrows and shifts to higher energy values in the case of the WSe$_2$ monolayer, as expected for a semiconductor. The same tendency is observed for the vdWH until the spin transition takes place and a considerable redshift and broadening of the peak occurs. The PL peaks can be fitted well by considering a Pearson IV distribution (see **Methods**), as already reported for other transition metal dichalcogenide monolayers.[32,33] The obtained PL maximum and broadening, Γ, are shown in **Figure 5.e** and **Figure 5.f**, respectively. For the reference WSe$_2$ monolayer, the thermal dependence of the PL position and broadening is well described by the Varshni and Rudin-Reinecke-Segall law[34,35] (see **Methods**), respectively, in all the temperature range. For the vdWH case, the PL maximum position and peak broadening exhibit a clear hysteric behavior when the spin transition takes place. The broadening and shift of the PL peak are in agreement with the calculated evolution of the band structure for a WSe$_2$ monolayer under compressive strain, where there is a crossover of the minimum of the conductance band from Γ to Λ.[36] In particular, from the PL measurements versus temperature (**Figure 5**), it is observed a band gap diminishment of *ca.* 40 meV when comparing the reference WSe$_2$ with the vdWH. This value corresponds to a biaxial compressive strain of around 3 %, as reported by Chang *et al.*[36] and matches the expected strain value from our SCO (Δb/b ~ 4.4 % and Δc/c ~ 2.5 %, where b and c are the in-plane lattice parameters; see **Figure 1.b** and **Supplementary Section 6**). Overall, similar trends are observed for various thermal cycles as well as in different vdWHs with the WSe$_2$ monolayer placed either above or below the SCO layer (see **Supplementary Section 5.3**). The broadening and shift of the PL peak is more significant when the WSe$_2$ monolayer is on top of the SCO thin-layer (SCO/WSe$_2$, vdWH type H), probably due to the additional formation of out-of-plane ripples.[37] In addition, to disentangle the possible proximity and pressure effects, we place a thin h-BN layer between the WSe$_2$ monolayer and the SCO thin-layer, thus forming the sequences WSe$_2$/h-BN/SCO (vdWH type I) and SCO/h-BN/WSe$_2$ (vdWH type J). In these cases, the analysis of the PL spectra is more complex since a significant inter-layer electron-phonon coupling occurs at WSe$_2$/h-BN interfaces.[38]

Nonetheless, the effects of the spin transition are noticeable although the overall effect is smaller than when no h-BN is placed (see **Supplementary Section 5.3**).

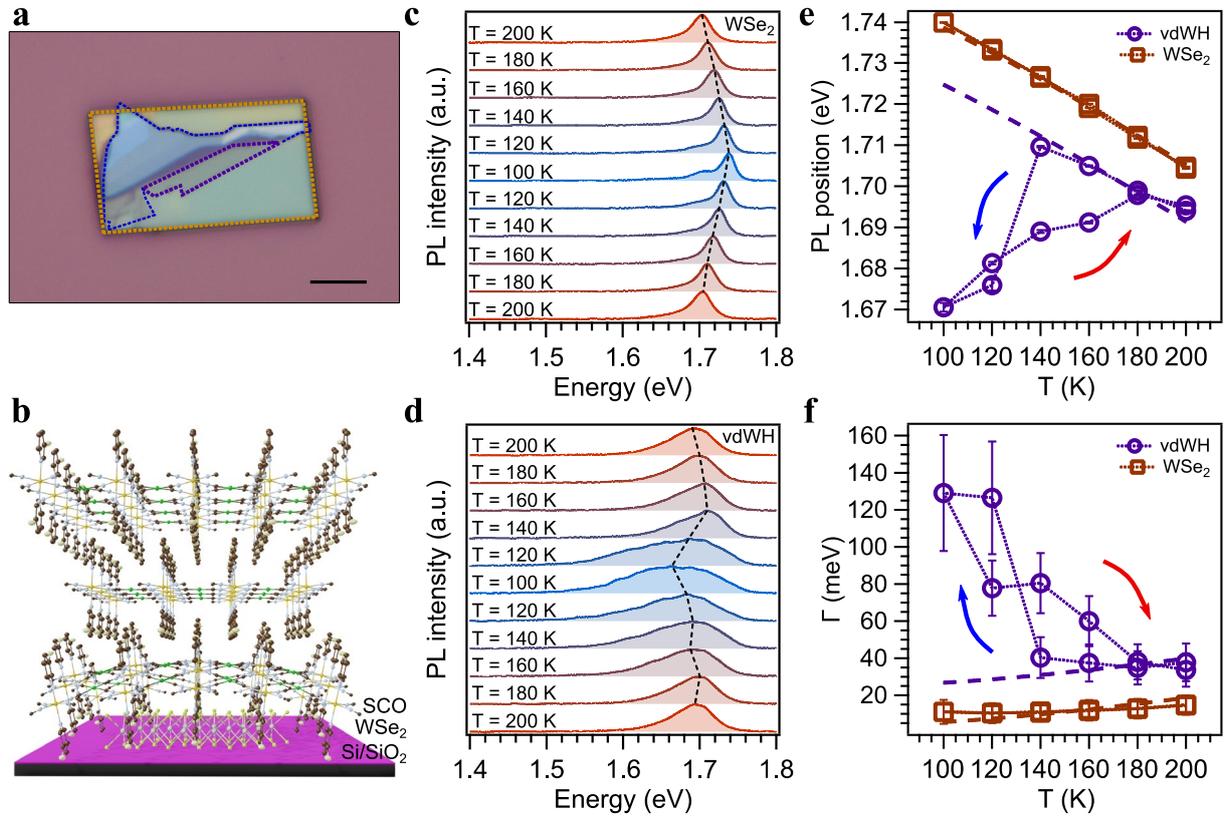

**Figure 5.-** Optical characterization of WSe$_2$/SCO vdWHs (third cycle for the vdWH G.1 in the **Supplementary Information**). a) vdWH formed by a WSe$_2$ monolayer (purple dashed lines) below a SCO thin-layer (orange dash lines). Scale bar: 5 μm. b) Artistic view of the vdWH. Atoms are represented as black (carbon), green (chlorine), orange (iron), blue(nitrogen), grey (platinum), yellow (selenium) and dark grey (tungsten) balls. c-d) Normalized photoluminescence at different temperatures for a reference WSe$_2$ monolayer (c) and the vdWH (d). For clarity, a vertical offset is added. Black dashed lines denote the maximum intensity position for the different spectra. d) Thermal dependence of the PL position for the vdWH and a reference WSe$_2$ monolayer. Fits following the Varshni law are marked as dashed lines (for the vdWH, it is only considered the points at high spin). e) Thermal dependence of the line broadening for the vdWH and a WSe$_2$ monolayer. Fits following the Rudin-Reinecke-Segall law are marked as dashed lines (for the vdWH, it is only considered the points at high spin). The cooling/warming direction is denoted with a blue/red arrow.

## 3. Conclusion

In summary, we have shown that the spin transition phenomenon occurring in molecular spin-crossover layers can switch the electrical and optical properties of 2D materials when assembled in van der Waals heterostructures. In particular, the electrical conductivity of FLG and the WSe$_2$-photoluminescence have been clearly modified by a thermal spin transition. This is attributed to the strain generated by the molecular layers due to the volume change concomitant with the spin transition. In the case of FLG this strain shifts the Fermi level, while in WSe$_2$ monolayers it changes the band structure. The present work incorporates spin-crossover layers as a novel building block able to induce strain in van der Waals heterostructures. Since the spin transition can be tuned by chemical design and triggered by a variety of physical stimuli, these molecular spin-crossover layers offer new routes for band engineering in 2D materials.

## 4. Experimental Section

*Crystal growth:* [Fe(3-Clpy)$_2$][Pt(CN)$_2$]$_2$ is synthesized by adapting the procedure described in reference [23]. High quality crystals of NbSe$_2$ are grown by Chemical Vapor Transport (CVT) using iodine as a transport agent, as already reported by some of us.[39] Further characterization can be found in the **Supplementary Section 1.**

*vdWHs fabrication:* Bulk crystals of SCO, NbSe$_2$ and natural graphite are exfoliated mechanically by the so-called *'Scotch-tape'* method and placed on top of 285 nm SiO$_2$/Si substrates using an adhesive tape (80 μm thick adhesive plastic film from Ultron Systems). As a prior identification, the obtained flakes are examined by optical microscopy (NIKON Eclipse LV-100 optical microscope under normal incidence). Atomic force microscopy images are taken with a Nano-Observer AFM from CSI Instruments. The vdWHs are placed on top of pre-lithographed electrodes (5 nm Ti/50 nm Au on 285 nm SiO$_2$/Si from NOVA Electronic Materials, LCC) by the deterministic assembly of the flakes using polycarbonate, as reported in reference [40], using of a micromanipulator. The whole process is performed inside inert atmosphere conditions (argon glovebox). Regarding the Hall bar devices, mechanically exfoliated graphite flakes are deposited on SiO$_2$/Si substrates and inspected in an optical microscope. The samples are then spin coated with Poly(methyl methacrylate) (PMMA 495 kDa) and baked at 180 °C for 7 min. In order to shape the Hall bar configuration, the areas surrounding the Hall bar are exposed with conventional electron-beam lithography, and etched with $O_2$ plasma (exposed 20 s with a pressure of 0.05 mbar and a power of 25 W) in a plasma reactive-ion etching equipment (ICP Etch System SI500 from Sentech). The remaining resist is removed with an acetone bath at room temperature for 3 min, rinsed with isopropanol and dried with a nitrogen gun. For the electrical contacts of the Hall bar, the sample is spin coated with a double-layer PPMA (495/950 kDa) and the metal contacts are fabricated by electron beam lithography techniques (5 nm Ti/ 50 nm Au). The lift-off is performed with acetone at room temperature.

*Optical contrast measurements:* Optical measurements are recorded with a camera (DCC1645C-HQ from Thorlabs) connected to an optical tube and a 20x objective. Temperature sweeps are performed with an open-flow nitrogen cryostat (manufactured by CryoVac), with top optical access (fused silica AR-coated window). The contrast is calculated as $C = (I_{SCO} - I_{substrate})/(I_{SCO} + I_{substrate})$, being $I_{SCO}$ the intensity of the SCO and $I_{substrate}$ the intensity of the substrate.[41] Further details are given in the **Supplementary Section 3**.

*Electrical characterization:* Electrical measurements are performed in a Quantum Design PPMS-9 cryostat with a 4-probe geometry, where a DC current is passed by the outer leads and the DC voltage drop is measured in the inner ones. DC voltages and DC currents are measured (MFLI from Zurich Instruments) using an external resistance of 1 MΩ, i.e., a resistance much larger than the sample.[41] Temperature sweeps are performed at 1K/min. In particular, for assessing the reproducibility over different thermal cycles, we consider the transition temperatures for cooling ($T_c^\downarrow$) and warming ($T_c^\uparrow$) –defined as the point with the maximum slope change–, the hysteresis height and the activation energies involved above and below the transition temperatures considering an Arrhenius law, where the conductance, G, is modelled as $G = G_0 \cdot \exp(-E_a/k_B T)$, being $G_0$ a prefactor, $E_a$ the activation energy, $k_B$ the Boltzmann constant and T the temperature. We define ΔR as the difference of the resistance mean value within the hysteresis loop between the warming and cooling sweeps. As well, for taking into account possible inaccuracies in the geometrical factors of the different devices, we normalized the resistance, $R_{norm}$, as $R_{norm}(T) = 100*[R(T) - R(T^*)]/R(T^*)$, where $R(T^*)$ is the resistance at $T^*$, a temperature value out of the hysteretic region. In our case, we systematically consider $T^* = 185$ K although we note that the election of $T^*$ does not vary significantly the value of $\Delta R_{norm}$ (defined in an analogous way as $\Delta R$), see the **Supplementary Section 4.4**. As an example, in the case of the cycle shown in the **Figure 2**, the following values are obtained: $T_c^\downarrow = 140$ K, $T_c^\uparrow = 166$ K, $\Delta R = (159 \pm 5)$ Ω, $\Delta\rho = (13.6 \pm 0.4)$ μΩ·cm, $\Delta R_{norm} = (2.91 \pm 0.13)$ %, $E_a = (1.77 \pm 0.02)$ meV for the LS region and $E_a = (5.18 \pm 0.06)$ meV for the HS region.

*Photoluminescence measurements:* PL measurements are performed in a LabRam HR Evolution confocal Raman microscope (Horiba) using 0.50 NA 50x objectives (Olympus LMPlanFL N), an incident wavelength of 532 nm and a grating of 600 gr/mm. The incident power to the sample is 17.26 μw. Temperature is controlled with a Linkam THMS600 stage. The measurement protocol is as follows: once the temperature setpoint is reached, we wait 10 minutes for a better thermal stabilization and, then, we measure the PL for the vdWH and for the $WSe_2$ reference monolayer. The spectra for vdWH type G are fitted following a Pearson IV distribution, as already reported for $MoS_2$, $MoSe_2$ and $WSe_2$ monolayers[32,33] (see **Supplementary Section 5** for further details). The thermal dependence of the PL maximum, $E(T)$, is modelled by the Varshni law for a semiconductor[34] and the line broadening, $\Gamma(T)$, to the Rudin-Reinecke-Segall law,[35] respectively. For the former, $E(T) = E(0) - \frac{\alpha T^2}{T+\beta}$ while, for the latter, $\Gamma(T) = \Gamma(0) + \sigma T + \Gamma' \frac{1}{e^{\frac{\hbar\omega}{kT}}-1}$. In the previous expressions, $E(0)$ is the energy gap at 0 K, α and β are fitting parameters related to the temperature-dependent dilatation of the lattice and Debye's temperature, respectively, $\Gamma(0)$ is the broadening at 0 K, σ and Γ' reflects the interaction between excitons and acoustic and longitudinal optical phonons, $\hbar\omega$ is the longitudinal optical phonon energy and k is the Boltzmann constant.[42,43]

**Supporting Information**

Supporting Information is available from the Wiley Online Library or from the author.

**Acknowledgements**

We acknowledge the financial support from the European Union (ERC AdG Mol-2D 788222 and COST Action MOLSPIN CA15128), the Spanish MICINN (MAT2017-89993-R, RTI2018-098568-A-I and EQC2018-004888-P co-financed by FEDER and Excellence Unit "María de Maeztu", CEX2019-000919-M), and the Generalitat Valenciana (PO FEDER Program, ref. IDIFEDER/2018/061). C.B.-C. thanks the Generalitat Valenciana for a PhD fellowship. We thank Safiya Ouazza for her help regarding the $WSe_2$-based vdWHs and Ángel López-Muñoz for his constant technical support and helpful discussions.

**ToC**

Layered spin-crossover metal organic frameworks are deterministically assembled with other two-dimensional (2D) materials, as graphene or WSe$_2$, forming van der Waals heterostructures. The strain concomitant to the spin transition clearly switches the electrical and optical properties of the 2D material. Thus, spin-crossover van der Waals heterostructures represent a new route for band engineering in low-dimensional materials.

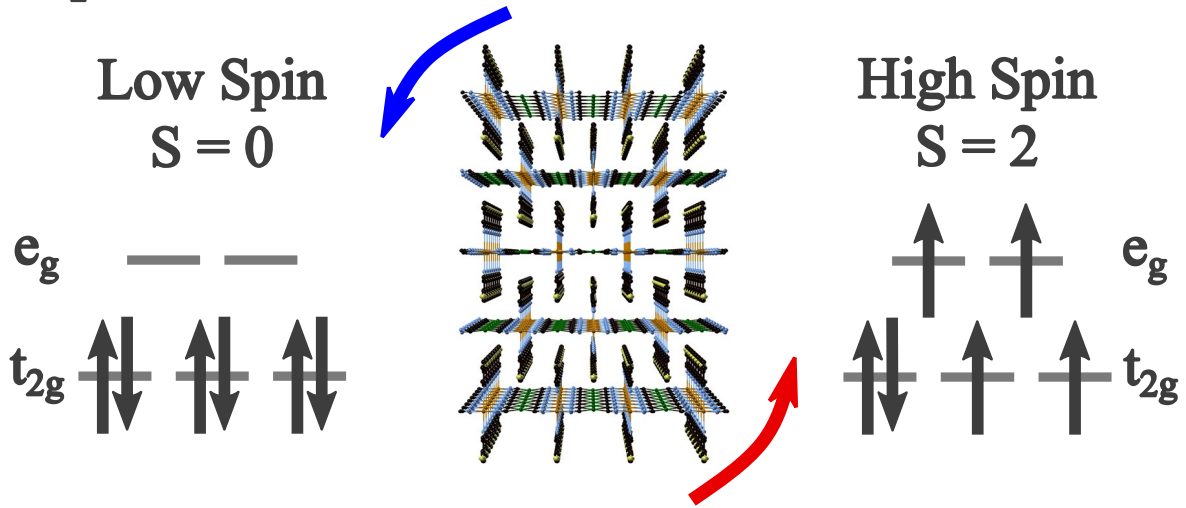

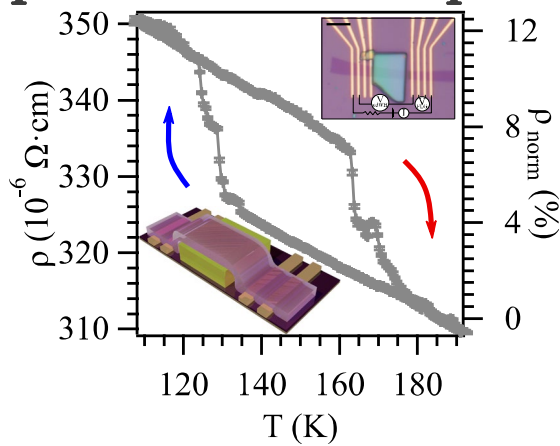

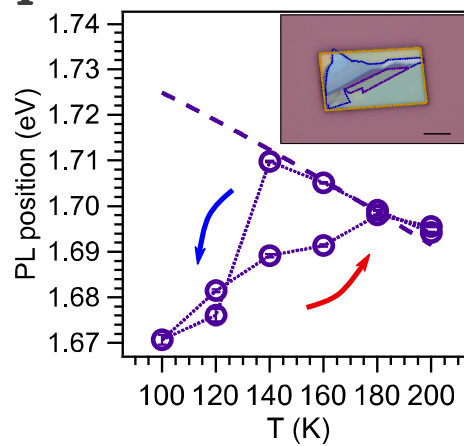